# TL-nvSRAM-CIM: Ultra-High-Density Three-Level ReRAM-Assisted Computing-in-nvSRAM with DC-Power Free Restore and Ternary MAC Operations

Dengfeng Wang[1], Liukai Xu[1], Songyuan Liu[1], Zhi Li[1], Yiming Chen[2], Weifeng He[1], Xueqing Li[2*], Yanan Sun[1*]

[1]Department of Micro-Nano Electronics, Shanghai Jiao Tong University, Shanghai 200240, China
[2]Department of Electronic Engineering, Tsinghua University, Beijing 100084, China
*Corresponding Authors: {sunyanan@sjtu.edu.cn, xueqingli@tsinghua.edu.cn}

## ABSTRACT

Accommodating all the weights on-chip for large-scale NNs remains a great challenge for SRAM based computing-in-memory (SRAM-CIM) with limited on-chip capacity. Previous non-volatile SRAM-CIM (nvSRAM-CIM) addresses this issue by integrating high-density single-level ReRAMs on the top of high-efficiency SRAM-CIM for weight storage to eliminate the off-chip memory access. However, previous SL-nvSRAM-CIM suffers from poor scalability for an increased number of SL-ReRAMs and limited computing efficiency. To overcome these challenges, this work proposes an ultra-high-density three-level ReRAMs-assisted computing-in-nonvolatile-SRAM (TL-nvSRAM-CIM) scheme for large NN models. The clustered n-selector-n-ReRAM (cluster-nSnRs) is employed for reliable weight-restore with eliminated DC power. Furthermore, a ternary SRAM-CIM mechanism with differential computing scheme is proposed for energy-efficient ternary MAC operations while preserving high NN accuracy. The proposed TL-nvSRAM-CIM achieves 7.8x higher storage density, compared with the state-of-art works. Moreover, TL-nvSRAM-CIM shows up to 2.9x and 2.0x enhanced energy efficiency, respectively, compared to the baseline designs of SRAM-CIM and ReRAM-CIM, respectively.

## 1 INTRODUCTION

With the rapid evolution of deep neural networks (DNNs), the parameter volumes of the state-of-art NN models are explosively growing and getting deeper. Properly handling the enormous number of parameters and computations in an energy-efficient manner is pivotal to prevail artificial intelligence (AI) in various applications, especially on edge devices with limited hardware resources. Computing-in-memory is a promising computing paradigm to accelerate NNs by transcending the memory-wall in the traditional von Neumann architecture.

The Static Random-Access Memory (SRAM) [1]-[3] and non-volatile Resistive Random-Access Memory (ReRAM) [4]-[8] are two attractive candidates for realizing CIM. Meanwhile, the foundry-mature SRAM-CIM array possesses high computation speed and energy efficiency, but at the cost of large area with quite limited capacity of on-chip volatile SRAMs. The data transfer between the SRAM-CIM memory and the off-chip weight-storing memory is thus still inevitable and costly for accelerating large-scale NNs with SRAM-CIM as shown in Fig. 1(a). The non-volatile ReRAM-CIM with single-level cells (SLCs) [4]-[5] or multi-level cells (MLCs) [6]-[7] enables higher weight-storage density with a smaller device footprint than SRAM-CIM, as shown in Fig. 1(b). However, ReRAM-CIM is sensitive to stochastic resistance variations, which results in error-prone multiply-and-accumulate (MAC) results and accuracy loss. Plenty of efforts have been reported in recent works, such as write-and-verify programming methods [7]-[8], unary weight-coding [9], and error-recovery training enabled by on-chip SRAM block [10]. However, the accuracy loss still cannot be fully recovered, which remains a severe challenge to employ high-robustness large-scale NNs with ReRAM-CIM.

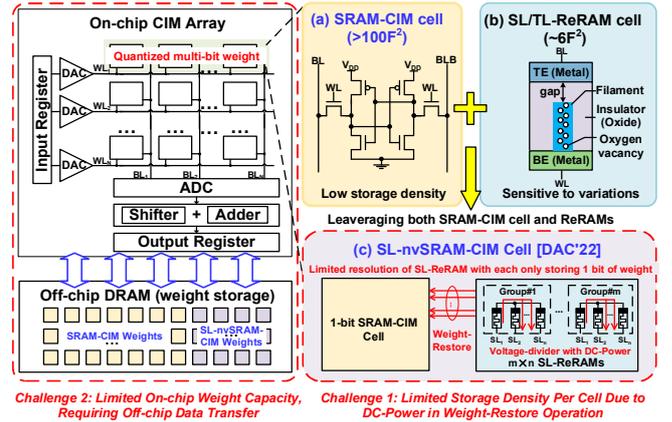

Figure 1: Different CIM techniques with design challenges. (a) SRAM-CIM [1]-[3]. (b) ReRAM-CIM [4]-[7]. (c) SL-nvSRAM-CIM [12].

To achieve high-density weight capacity while maintaining the accuracy of CIM, hybrid CIM architectures [11]-[12] combining nonvolatile memory and SRAM-CIM have been presented. In [11], the STT-MRAMs are used for on-chip weight storage to alleviate the off-chip memory access problem in SRAM-CIM. However, [11] remains an isolated computing system with limited on-chip bandwidth for energy-cost data transfer. In [12], computing-in-nonvolatile-SRAM with single-level ReRAMs (SL-nvSRAM-CIM) has been presented as an alternative solution for enlarged weight capacity on-chip where multiple SL-ReRAMs storing binary weights are attached to each bit of nvSRAM-CIM cell, as shown in Fig. 1(c). The multi-bit weights are restored to SRAM-CIM cells in array-level parallelism for MAC operations by utilizing the voltage-divider-type select scheme. However, the voltage-divider-type select scheme induces DC power in the SL-nvSRAM-CIM cell during weight-restore, degrading the reliability. This is even worse with an increased number of SL-ReRAMs. Consequently, the maximum allowed number of SL-ReRAMs is stringently constrained, which limits the storage density in each SL-nvSRAM-CIM cell. Furthermore, as each SLC can only store 1-bit of quantized weight, the overall on-chip weight capacity of SL-nvSRAM-CIM is limited and additional off-chip weight reloading is unavoidable for larger NN models, which hinders the efficient acceleration of DNNs in [12].

In this work, an ultra-high-density and energy-efficient multi-level ReRAM-assisted computing-in-nonvolatile-SRAM (TL-nvSRAM-CIM) is proposed for accelerating large NNs on-chip. The DC-power-free weight-restore mechanism is developed which enables ultra-high-density three-level ReRAMs (TL-ReRAMs) on-chip. High-efficiency CIM operations are supported for multiplying and accumulating the ternary weights and inputs with reduced computing cycles compared to binary-based CIM operations. The primary contributions of this work are concluded as follows:

(1) An ultra-high-density TL-nvSRAM-CIM cell with DC-power-free weight-restore is proposed to significantly enhance the storage capacity. The multi-trit ternary weights stored in clusters of three-level MLCs can be reliably restored to every pair of two neighboring SRAM cells with array-level parallelism. The clustered nSnRs are employed for reliable weight-restore while eliminating the DC power, despite of MLC variations. The number of TL-ReRAMs can be therefore significantly enlarged with high restoring reliability. The proposed TL-nvSRAM-CIM arrays enhance the storage density by 7.8x as compared to the previous SL-nvSRAM-CIM arrays.

(2) A ternary SRAM-CIM mechanism with differential computing scheme is proposed to enhance the energy efficiency while preserving the robustness of the TL-nvSRAM-CIM circuit for NN acceleration. Multiple trits of ternary weights and inputs can be directly multiplied and accumulated by using the differential computing paths in multiple pairs of neighboring SRAM cells of TL-nvSRAM-CIM circuit.

(3) Evaluation results show that the proposed TL-nvSRAM-CIM architecture is adaptive to larger-scale NN models compared to the previous SL-nvSRAM-CIM. TL-nvSRAM-CIM shows up to 2.9x and 2.0x energy efficiency compared to SRAM-CIM and ReRAM-CIM, respectively, and achieves 11.0x energy efficiency per unit area compared to SL-nvSRAM-CIM.

The remainder of the paper is organized as follows. In Section 2, the challenges of the previous CIM circuits are introduced. The proposed TL-nvSRAM-CIM for NN acceleration is presented in Section 3. The characteristics of the proposed TL-nvSRAM-CIM and traditional CIM are evaluated and further discussed in Section 4. Finally, conclusions are provided in Section 5.

## 2 CHALLENGES OF PREVIOUS CIM

The background and challenges of previous in-memory computing architectures are presented in this section. The previous SRAM- and ReRAM-CIM works are reviewed in Section 2.1. The previous SL-nvSRAM-CIM work is reviewed in Section 2.2.

### 2.1 Previous SRAM and ReRAM based CIM

The current-domain [1]-[2] and charge-domain [3] SRAM-CIMs have gained a lot of attentions in recent years. The current-domain SRAM-CIM utilizing the accumulated current produced on bitlines for in-memory MAC is featured with simplified circuit structure with relatively small area cost [1]-[2]. While the charge-domain

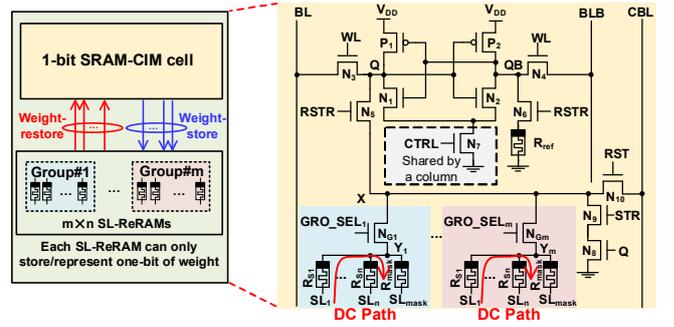

Figure 2: Schematic of previously presented SL-nvSRAM-CIM cell [12] which stores and restores only one-bit weight.

SRAM-CIM has more robust and accurate MAC results due to less capacitor mismatch [3]. However, given the constrained resources of edge devices, SRAM cannot hold all the weights for large NN computing. The data transfer between SRAM and off-chip DRAM is still required for reloading the weights which deteriorates the energy efficiency of SRAM-CIM at system level.

With the footprint as small as $6F^2$, ReRAM-CIM based on SLCs [4]-[5] or MLCs [6]-[7] enables high-density weight storage. However, the MAC results of ReRAM-CIM are sensitive to device variations leading to significantly degraded accuracy of NNs which could become even worse for MLC with squeezed neighboring resistance states. To alleviate the accuracy loss in ReRAM-CIM, several works are presented in [7]-[10]. In [7]-[8], the write-and-verify programming method is used to reduce the deviations of ReRAM resistance during write. The variation-aware algorithms named unary coding [9] could alleviate the variations of synaptic weights. On-chip training method [10] using SRAM block with trainable weights could recover the accuracy loss caused by the ReRAM variations. However, the accuracy of ReRAM-CIM cannot be fully regained. It remains a great challenge to realize the high accurate ReRAM-CIM.

### 2.2 Previous SL-NvSRAM-CIM

To achieve high on-chip weight capacity while preserving high NN accuracy, the SL-nvSRAM-CIM is presented in [12] by employing groups of SL-ReRAMs for storing multi-bit weights, as shown in Fig. 2. The high-resistance state (HRS) and low-resistance state (LRS) of each SL-ReRAM are used to represent binary logic '0' and '1', respectively. By using the voltage-divider-type select scheme, each bit of weights can be selectively restored to SRAM-CIM cell for binary MAC operation. During weight-restore operation, the selected SL-ReRAM ($R_{s1}$ in Fig. 2) is serially cascaded with the parallelly connected unselected SL-ReRAMs with a total low resistance, which forms the voltage divider. The states of selected SL-ReRAM can be distinguished by the output voltages of voltage divider. However, the voltage-divider-type select scheme induces DC in each ReRAM groups with degraded weight-restore reliability. As the number of SL-ReRAM increases in a voltage divider, the two divided voltages representing the HRS and LRS become closer and less likely to be resolved due to the increased combinations of unselected ReRAM resistances. Given CMOS and ReRAM variations, the selected SL-ReRAMs cannot be restored to SRAM-CIM cells correctly which severely limits the maximum allowable number of ReRAMs per SL-nvSRAM-CIM cell. Furthermore, SL-ReRAM can only store 1-bit weight for NN. Consequently, the on-chip weight capacity of SL-nvSRAM-CIM is inherently limited. Given that the large NN is employed in SL-nvSRAM-CIM, additional data transfer between on-chip SRAM-

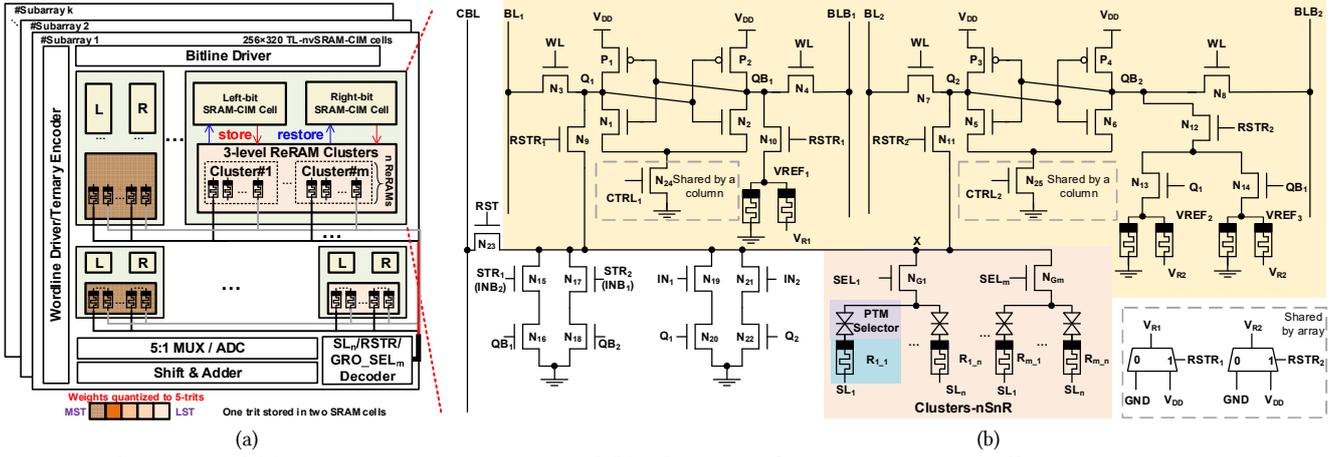

Figure 3: The proposed (a) TL-nvSRAM-CIM macro and (b) schematic of TL-nvSRAM-CIM cell.

Table 1: Differential signals and ReRAM values representing ternary inputs and weights.

| Coding Value | Input (each trit of input) | | |
|---|---|---|---|
| Ternary trit value | +1 | 0 | -1 |
| $IN_1/IN_2$ ($INB_1/INB_2$) | 1/1 (0/0) | 1/0 (0/1) | 0/0 (1/1) |
| Coding Value | Weight (each trit stored in 3-level-ReRAM) | | |
| Ternary trit value | +1 | 0 | -1 |
| $Q_1/Q_2$ (ReRAM state) | 0/0 (LRS) | 1/0 (MRS) | 1/1 (HRS) |

Table 2: Signal settings of TL-nvSRAM-CIM circuit.

| Signals | | $SEL_i$ | $SL_j$ | $SL_x$ | $RSTR_1$/$RSTR_2$ | $STR_1$ | $STR_2$ | CBL |
|---|---|---|---|---|---|---|---|---|
| Parallelism | | Array | | | | Row | | Column |
| Store Mode | Phase 1 | $V_{DDH}$ | GND | $V_{DDL}$ | GND | GND | GND | $V_{DDH}$ |
| | Phase 2 | $V_{DDH}$ | $V_{DDH}$ | $V_{DDL}$ | GND | $V_{DD}$ | $V_{STR}$ | Floating |
| Restore Mode | Phase 1 | GND | $V_{DDL}$ | $V_{DDL}$ | GND | GND | GND | Floating |
| | Phase 2 | $V_{DD}$ | GND | $V_{DDL}$ | $V_{DD}$ | GND | GND | Floating |
| CIM Mode | | GND | $V_{DDL}$ | $V_{DDL}$ | GND | $INB_2$ | $INB_1$ | MAC |

*$SEL_i$ and $SL_j$: Signals connected to the selected TL-ReRAMs; $SL_x$: Source line connected to unselected TL-ReRAMs; $V_{DD}$=0.9V, $V_{DDH}$=1.5V, $V_{DDL}$=0.6V, $V_{STR}$=0.31V.

CIM and off-chip memory is still required which significantly impairs the energy efficiency.

## 3 PROPOSED TL-NVSRAM-CIM

To achieve ultra-high-density with reliable weight-restore, this section presents a new TL-nvSRAM-CIM macro, as illustrated in Fig. 3(a). The TL-nvSRAM-CIM macro quantizes NN weights to multiple trits which are stored in TL-ReRAMs for significantly enhanced storage density and restored to pairs of neighboring SRAM cells without DC-power for energy-efficient ternary MAC operations.

### 3.1 Overview of Proposed TL-nvSRAM-CIM

As shown in Fig. 3(a), to perform the acceleration of large-scale NN, the proposed TL-nvSRAM-CIM scheme comprises six subarrays with 256×320-bits SRAM cells and peripheral circuits, including ternary input driver, MUX, ADC, shift & adder, and array signal controller. Every two neighboring SRAMs on the same row and corresponding ReRAM clusters make up an TL-nvSRAM-CIM cell. The capacity of the weights stored in ReRAMs are significantly magnified by the TL-ReRAMs to accommodate large-scale NNs. Meanwhile, by storing weights in TL-ReRAMs, the circuit can be powered off for saving energy when the accelerator is not working.

The NN weights are quantized to multi-trit and stored in the selected TL-ReRAMs sharing the same source line (*SL*) from a certain cluster in subarray. Before the NN inference, each trit of weights is restored from one TL-ReRAM to two neighboring SRAM cells in array-level parallelism. The 8-bit input is encoded to 5-trit by the ternary input driver, which is shared by 16 rows with negligible area and energy cost. Each trit of weight and input is represented by two signals $Q_1Q_2$ and $IN_1IN_2$, respectively, as shown in Table 1. With 16 rows activated at the same time, the MAC operation is performed based on the ternary SRAM-CIM mechanism and differential computing scheme. The output on each bitline shared by every two SRAM columns is sensed by the ADC and the MAC value is generated by the shift & adder. Weights in different TL-ReRAMs are restored to SRAMs in different cycles, and multiple cycles are needed for completing the whole NN. The energy consumption and latency induced by the restore operation can be ignored, thanks to the DC-free and array-level parallel restore operations.

### 3.2 Proposed TL-nvSRAM-CIM Circuit

The schematic of proposed TL-nvSRAM-CIM cell is shown in Fig. 3(b). The TL-nvSRAM-CIM cell is composed of two 6T SRAM cells ($P_1$-$P_4$ and $N_1$-$N_8$), data restore paths ($N_9$-$N_{14}$), differential computing paths ($N_{15}$-$N_{23}$) which are partly shared with data store path ($N_{15}$-$N_{18}$), and cluster-nSnRs. The control transistors $N_{24}$ and $N_{25}$ are shared by columns during weight-restore. Three reference generators composed of serially connected ReRAMs are used to generate reference voltages ($V_{REF1}$/$V_{REF2}$/$V_{REF3}$). In this work, by using bidirectional selectors with excellent endurance [13], reliable DC-power-free weight-restore is realized. The insulator-metal transition voltage ($V_{IMT}$) and the metal-insulator transition voltage ($V_{MIT}$) of selectors are set as 0.45V and 25mV [13], respectively. The metallic-state resistance and insulating-state resistance of selectors are assumed to be 40kΩ and 0.12GΩ, respectively [17]. The LRS and HRS of three-level ReRAM in this work are assumed to be 80kΩ and 1MΩ, respectively [18]. The medium-resistance state (MRS) is determined by maximizing the minimum value of MRS/LRS and HRS/MRS [14] for high restoring reliability, which is evaluated as 282kΩ. The balanced-ternary coded weights are stored in TL-ReRAMs with HRS for '-1', MRS for '0', and LRS for '+1'.

Three modes are supported by the proposed TL-nvSRAM-CIM circuitry: store mode (storing each trit of weights from one pair of SRAM cells to TL-ReRAM devices in array-level parallelism), restore mode (backup the weights in TL-ReRAM to pairs of SRAM-CIM cells in array-level parallelism), and CIM mode (ternary MAC operations). The important signal settings of TL-nvSRAM-CIM with ternary inputs and weights are listed in Table 2.

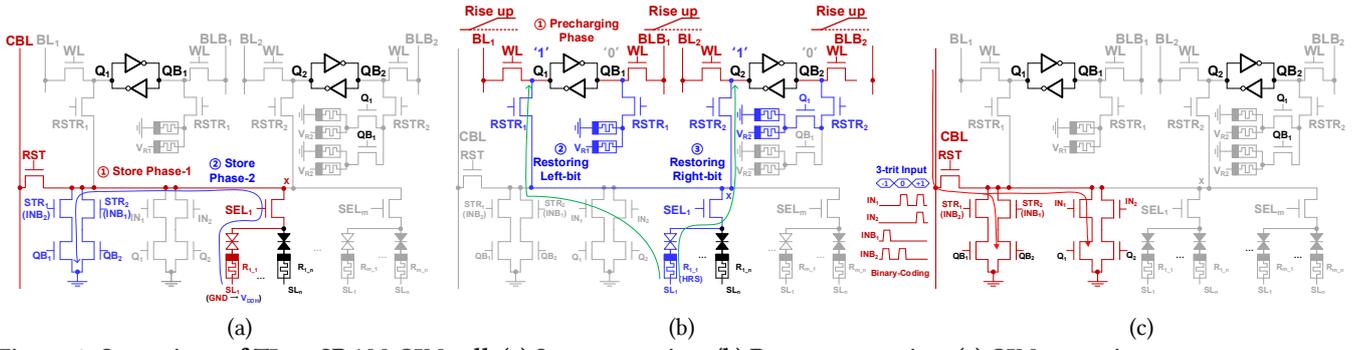

Figure 4: Operations of TL-nvSRAM-CIM cell. (a) Store operation. (b) Restore operation. (c) CIM operation.

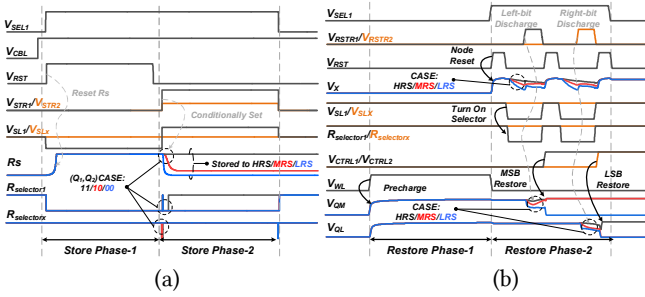

Figure 5: Simulated waveforms of (a) store operation and (b) restore operation.

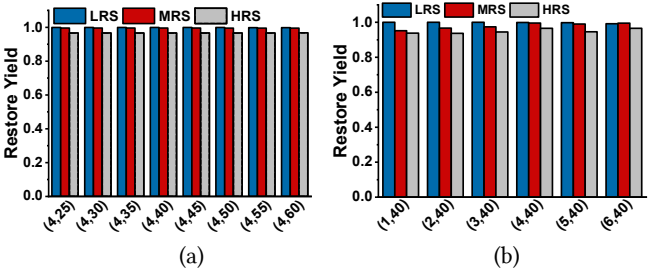

Figure 6: Data restore yield considering both ReRAM and CMOS variations with (a) different ReRAM numbers $n$ and (b) different numbers of clusters $m$.

## 3.3 Weight-Store Operation

In the store mode, TL-ReRAMs are initialized with multi-trit weights in array-level parallelism. The store mode is composed of two phases as shown in Fig. 4(a). In store phase-1, the $SL_j$ of selected ReRAMs $R_{i\_j}$ (in the $i^{th}$ cluster and connected to the $j^{th}$ SL) pulls from $V_{DDL}$ to GND to turn on the selector while the $SLs$ of other unselected TL-ReRAMs maintains $V_{DDL}$. All the selectors of unselected TL-ReRAMs are in insulating states with no DC-power consumptions compared to the voltage-divider select scheme [12]. Then $RST$ and $SEL_i$ transitions to $V_{DDH}$ to create a reset path from $CBL$ to $SL_j$. All the selected $R_{i\_j}$ in the whole array are reset to HRS. In store phase-2, all the selected $R_{i\_j}$ are conditionally set depending on the value of storage nodes $Q_1$ and $Q_2$. The $SL_j$ of selected TL-ReRAM transitions to $V_{DDH}$ while the unselected $SLs$ maintain at $V_{DDL}$. The $STR_1$ and $STR_2$ are applied to different voltages to create different strengths of set paths. If '00' are stored in '$Q_1Q_2$', set current $I_{00}$ is produced from paths $N_{15}$-$N_{16}$ and $N_{17}$-$N_{18}$ to set the selected TL-ReRAM to LRS ('+1') as listed in Table 1. If '10' are stored in '$Q_1Q_2$', set current $I_{10}$ is produced from path $N_{15}$-$N_{16}$ to set the selected TL-ReRAM to MRS ('0'). If '11' are stored in '$Q_1Q_2$', no set current is produced thus the selected TL-ReRAM stays at HRS ('-1').

The simulated waveforms of store operations for HRS/MRS/LRS are shown in Fig. 5(a). The selected selectors are opened in store phase-2 only when the set paths are activated. Although the unselected selectors may experience a sudden on-off switch (the glitch in waveform) caused by the transition process of $SL_j$, it has little impact on the state of the unselected ReRAMs (the states of unselected ReRAM are not changed). To provide enough drivability in store operations, the driven voltages of the core transistors used in TL-nvSRAM-CIM are set to 1.5V without degrading the reliability of devices [20][25]-[26]. However, it is hard to precisely control the writing process of TL-ReRAMs and the TL-ReRAM may suffer from wide distributed resistance, possibly introducing error data in SRAM and degrading the inference accuracy. The impact of ReRAM variations on the robustness of proposed TL-nvSRAM-CIM is evaluated in Sections 3.4 and 4.3.

## 3.4 DC-Power-Free Weight-Restore Operation

In the restore mode, ternary weights stored in TL-ReRAMs are reloaded into the pairs of neighboring SRAM cells in array-level parallelism. The restore mode contains two phases that are precharging phase (restore phase-1) and sequentially restoring phase (restore phase-2) as shown in Fig. 4(b). In the precharging phase, the $CTRL_1$ and $CTRL_2$ are turned off and bitlines are precharged to $V_{DD}$. $WL$ transitions high to initialize storage nodes of pairs of SRAM cells to $V_{DD}$.

In the beginning of sequentially restoring phase, the left-bit SRAM is firstly restored. The $SL_j$ of $R_{i\_j}$ transitions from $V_{DDL}$ to GND to turn on the selector while the selectors of other unselected TL-ReRAMs remain at insulating state by setting the unselected $SLs$ to $V_{DDL}$. Note that, there is no DC-power induced when restoring $R_{i\_j}$ to SRAMs in the proposed cluster-nSnRs, compared to the voltage-divider select scheme in the previous SL-nvSRAM [12]. $Q_1$ and $QB_1$ start to differentially discharge by turning on $RSTR_1$ and $SEL_i$. $Q_1$ discharges through $R_{i\_j}$ while $QB_1$ discharges through reference voltage $V_{REF1}$ to generate enough voltage difference. Assuming $R_{i\_j}$ is HRS ('-1') or MRS ('0') inducing a lower discharge current in $Q_1$, $Q_1$ and $QB_1$ will be amplified to $V_{DD}$ and 0V, respectively, after turning on $CTRL_1$. If $R_{i\_j}$ is LRS ('+1') inducing a larger discharge current in $Q_1$, $Q_1$ and $QB_1$ will be amplified to 0V and $V_{DD}$, respectively, after turning on $CTRL_1$.

To restore the right-bit SRAM, $RSTR_2$ transitions high after the left-bit is restored. $Q_2$ and $QB_2$ start to differentially discharge. $Q_2$ is discharged through $R_{i\_j}$, while $QB_2$ discharges depending on the value of $Q_1$. When $Q_1$ is restored to '1', the $QB_2$ discharges through $V_{REF2}$ otherwise $V_{REF3}$. If $R_{i\_j}$ is HRS ('-1') (or MRS ('0')), the discharge path of $QB_2$ through $V_{REF2}$ is selected and generating a relatively smaller (or larger) discharge current in $Q_2$ compared to $QB_2$. When sufficient voltage difference is established between $Q_2$ and $QB_2$, $CTRL_2$ is turned on to amplify $Q_2$ and $QB_2$ to full swing. When $Q_1$ is restored to '0', the discharge path of $QB_2$ is selected through $V_{REF3}$, a larger discharge current is generated in $Q_2$ compared to $QB_2$. After $CTRL_2$ is turned on, $Q_2$ and $QB_2$ are amplified to 0V and $V_{DD}$, respectively.

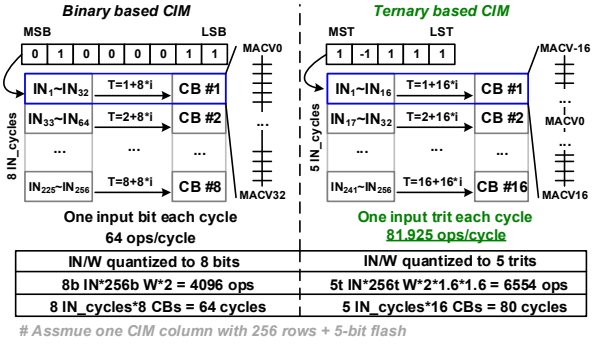

(a)

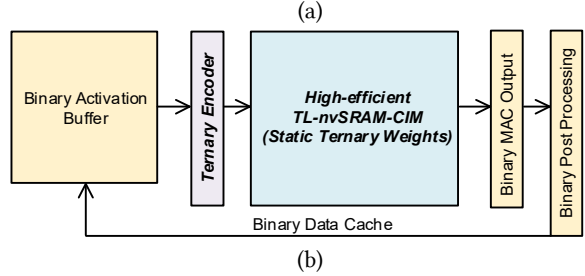

(b)

Figure 7: (a) Comparisons between binary coding and ternary coding based CIM. (b) The proposed high-efficiency ternary based computing flow which is fully compatible with binary computing system.

Table 3: The software inference accuracies of different coding methods on Cifar-10.

| Quantization Methods | ResNet-18 [23] | VGG-9 [24] |
|---|---|---|
| Floating Point Baseline | 96.5% | 94.5% |
| BC (8b-IN/8b-W) | 96.1% | 94.3% |
| TC (5t-IN/5t-W) | 95.4% | 94.2% |
| BC (8b-IN/8b-W) truncated to TC (5t-IN/5t-W) | 96.1% | 94.3% |

The simulated waveforms of restore operations for different cases (HRS/MRS/LRS) are shown in Fig. 5(b). To further evaluate impact of the discharge mismatch caused by CMOS and ReRAM variations, restore data yield (defined as the probability of successfully restoring TL-ReRAM to SRAMs) is measured through 1000 Monte-Carlo simulations considering both CMOS and ReRAM variations. The ReRAM variations which are considered as the filament gap variations 3σ/μ are assumed to be 10% [19][27]. The simulated results of different ReRAM settings (across various number of clusters or ReRAMs per cluster) are shown in Fig. 6. As contrast to the SL-nvSRAM-CIM which has limited number of ReRAMs in a single group [12], the proposed TL-nvSRAM-CIM maintains its data yield well above 94% even when the number of ReRAMs in a single cluster increases to 60. There are two key observations. Firstly, the exact resistance value of TL-ReRAM is less cared about in the TL-nvSRAM-CIM scheme as compared to traditional ReRAM-CIM (due to the possible accumulated errors). Since the TL-ReRAMs are digitalized by restore operations, what matters is whether the sensing margin between neighboring states is enough to be distinguishable. Secondly, TL-nvSRAM-CIM provides relatively large sensing margins (see $V_X$ in Fig. 5(b)) between different resistance states by employing selectors instead of the select scheme with narrowed divided voltages in [12].

Besides, the yield of different data patterns can be tuned by designing the discharge reference paths, considering the specific data distributions in task. In this work, the MRS is tuned as preference since weights in NN are sparse.

### 3.5 Ternary based CIM Operation

In CIM mode, multiplication between one-trit input and one-trit weight each cycle is realized by utilizing the differential CIM paths, as shown in Fig. 4(c). To perform the current domain SRAM-CIM, the $CBL$ is precharged to $V_{DD}$. If each trit of inputs is '+1' which is encoded as '11', the $IN_1$ and $IN_2$ are both applied with $V_{DD}$, while the complementary inputs $INB_1$ and $INB_2$ are set at GND. Different number of discharge paths will be activated, depending on the value of $Q_1$ and $Q_2$. When '$Q_1Q_2$' is '11', total two discharge currents are produced on $CBL$ which means multiplication result of '-1'. When '$Q_1Q_2$' is '10', only one discharge current is produced on $CBL$ which means multiplication result of '0'. When '$Q_1Q_2$' is '00', no discharge current is produced on $CBL$ which means multiplication result of '+1'. Other input cases (like '0' and '-1') are not described here for simplicity. Then the discharge currents representing one-trit input multiplying one-trit weight of each row are accumulated on $CBL$. With a fixed discharge time, the voltage of $CBL$ drops $ΔV$ proportional to the MACs.

The comparisons between binary based and ternary based CIM are shown in Fig. 7(a). In conventional binary based computing, inputs and weights are coded as binary bits (BC). In each computing cycle, the input slices are split into multiple sub-slices and sent to compute blocks (CBs). Constraint by ADCs, only 32 rows for one CB are activated at a time. The total computing cycles thus equals the product of input bit-width and number of CBs. The throughput of binary based computing is low due to the limited number of operations each cycle. In the proposed ternary based computing, the data width of inputs and weights is compressed by ternary coding (TC), resulting in less input cycles and higher number of operations each cycle. The throughput is therefore significantly enhanced in TC although the number of CB is increased due to the less activated rows in each cycle. Note that, conventional binary based computing serially sends multiple bits of one input to the CBs each cycle instead of one bit, which can be indicated as multi-bit-serial computing. In this scenario, the throughput is even degraded compared to the bit-serial one due to the less activated rows constrained by the ADCs. For example, for serializing 2-bit input each cycle, the activated rows are reduced to 11 instead of 32 which increases the CBs to 24 instead of 8 compared to bit-serial computing. The compute cycles are significantly enlarged while the total number of operations does not change, so the throughput is degraded in BC-based multi-bit-serial computing.

The proposed ternary based computing flow is shown in Fig. 7(b). Since the ternary coding is redundant for binary based digital systems, only the weights is statically stored in TL-nvSRAM-CIM array while the activations are dynamically converted to ternary with binary representations. The TL-nvSRAM-CIM array can output binary MAC results after ADC sensing without additional conversion. Therefore, the proposed ternary based computing flow can be fully compatible with existing binary computing systems for high-efficiency MAC operations.

The 256×320-bit TL-nvSRAM-CIM macro activates maximum 16 rows in this work for the enough voltage margins of MAC values. Voltages of $CBLs$ are fed into 5-bit ADC which is shared by every 5 TL-nvSRAM-CIM $CBLs$ to get the final results. Note that previous SL-nvSRAM-CIM [12] uses binary encoded inputs and weights for multi-bit MAC operations by serializing input bit every cycle with relatively large computing latency. In the proposed TL-nvSRAM-CIM, 8-bit inputs are encoded to 5-trit inputs, reducing the cycles of binary MAC operations in [12] from 8 to 5 with enhanced computing energy efficiency.

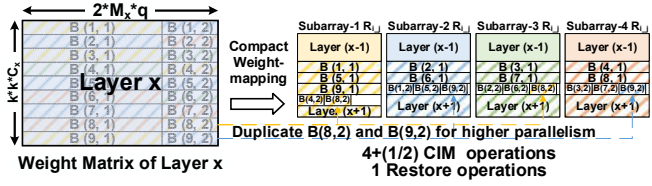

**Figure 8: An example of the compact weight mapping method for TL-nvSRAM-CIM scheme.**

The impact of different quantization methods on the inference accuracy are listed in Table 3. Since 8-bit quantization can present wider data range, directly quantitating weights to 5-trit instead of 8-bit may leads to accuracy loss. However, the weights of NNs are inherently sparse and most of the weights are confined to small values, we choose to firstly quantize the weights to 8-bit then truncate the values to 5-trit. The truncating based quantization method can achieve almost the same accuracy as 8-bit quantization.

### 3.6 Ternary Weight Mapping Method

To effectively exploit the large capacity of TL-nvSRAM-CIM and minimize the waste of hardware resources, a compact weight mapping method is proposed. The compact mapping method can be divided into three steps. Firstly, the weights of each NN layer are converted into a $R_L \times C_L$ weight matrix and split into weight blocks. A convolutional layer quantized to $q$ trits with $C$ input channels and $M$ output channels with $k \times k$ kernel size is converted into a $(C \times k \times k) \times (M \times q \times 2)$ weight matrix. The matrix is divided into several $R \times C$ blocks, where $R$ equals to the number of rows activated at the same time in CIM and $C$ refers to the total number of columns in the subarray. Then, the weight blocks are distributed to different subarrays evenly for high parallelism. Moreover, by duplicating weights and exploiting the idle subarrays, the parallelism of NN inference can be further enhanced, as shown in Fig. 8. Third, the blocks distributed to subarrays are mapped to TL-ReRAMs. As shown in Fig. 8, weight blocks with smaller sizes are prior to be mapped to empty columns left by the former block, to map weights compactly. When the $R_{i\_j}$ is filled, latter weight blocks are mapped to $R_{i\_(j+1)}$.

## 4 Evaluation

The proposed TL-nvSRAM-CIM cell and architecture are evaluated in this section. The experimental setup is described in Section 4.1. The characteristics of proposed TL-nvSRAM-CIM cell and the whole architecture are evaluated in Sections 4.2 and 4.3, respectively.

### 4.1 Experimental Setup

*Cell-level evaluations.* The electrical characteristics of the proposed TL-nvSRAM-CIM circuit is evaluated and compared with previously published SL-nvSRAM-CIM circuits. For fair comparison with the previous SL-nvSRAM-CIM cell, the layouts of SL- and TL-nvSRAM-CIM cells are all designed in 28nm technology with Cadence Virtuoso. The SPICE simulations with extracted layout parasitics are performed at 25°C, using the experimentally verified physics-based ReRAM [15] and selector Verilog-A model [16], respectively.

*Architecture-level evaluations.* To address the architecture-level throughput/energy efficiency benefits brought by the ultra-high density and high-efficiency MAC operations of TL-nvSRAM-CIM, four baselines are introduced as follows:

- Baseline-1 (small on-chip capacity but high-efficiency CIM): off-chip DRAM + SRAM-CIM, the NN weights are stored in off-chip DRAM and loaded to SRAM-CIM array for MAC operation.
- Baseline-2 (large but isolated on-chip memory, high-efficiency CIM): on-chip ReRAM + SRAM-CIM, the NN weights are stored in on-chip ReRAMs and loaded to SRAM in row-level parallelism for MAC operation by SRAM-CIM.
- Baseline-3 (large on-chip capacity but less efficient CIM): ReRAM-CIM, the weight storage and matrix-vector-multiplication are both performed by SL-ReRAM crossbars.
- Baseline-4 (moderate on-chip capacity with high-efficiency CIM): previous state-of-the-art SL-nvSRAM-CIM [DAC'22], each SL-nvSRAM-CIM cell is attached with 18 SL-ReRAMs (3 groups and 6 SL-ReRAMs per group), maintaining relatively high restore yield.

We adopt two typical NN models of ResNet-18 (relatively large, 11MB) [23] and VGG-9 (relatively small, 3MB) [24] on Cifar-10 dataset to evaluate the performance and adaptability of different CIM architectures. The array sizes of binary based CIM architectures (four baselines) are set to 256×256 with 8-b quantized inputs and weights. Every 8 columns of memory are assumed to share the same 5-bit ADC, with 32 rows activated at the same time to fully exploit the resolution of ADC. For fair comparison with binary CIM, the array size of ternary based TL-nvSRAM-CIM is set to 256×320 with same number of 5-t quantized weights/inputs. Each ADC is shared by every 5 $CBL$s corresponding to 10 columns of SRAMs in the TL-nvSRAM-CIM. Therefore, the number of ADCs between binary and ternary based CIM in a single array is the same (e.g., 32). The architecture-level design specifications are listed in Table 5.

The accuracy evaluations of SL- and TL-nvSRAM-CIM are done by injecting the bit errors (induced by incorrect restore operations) into weight matrix, followed with retraining. The restore yield is used to present the probability of bit errors, which is measured in 1000 SPICE Monte-Carlo simulations considering both CMOS (TT corner) and ReRAM (filament gap $3\sigma/\mu$=10%) variations [19][27].

To further illustrate the capacity/density benefits of the proposed TL-nvSRAM-CIM, extra baselines are used:

- SL-nvSRAM-CIM with selectors: Since the DC-power-free select scheme can also be adopted to SL-nvSRAM-CIM, the number of ReRAMs contained in a single group can be increased to 18 (constraint by physical layout) with 3 groups in total.
- SL-nvSRAM-CIM with the same capacity: The number of SL-nvSRAM-CIM arrays should be enough to accommodate the largest model (e.g. ResNet-18, 11MB) used in the architecture-level evaluations.
- SL-nvSRAM-CIM with the same area: The number of SL-nvSRAM-CIM arrays should be constraint so as to maintain the similar area overhead compared with the proposed high-density TL-nvSRAM-CIM.

### 4.2 TL-nvSRAM-CIM Cell

The detailed design metrics of our proposed TL-nvSRAM-CIM circuit are demonstrated in Table 4. By employing the cluster-

**Table 4: Comparison between the proposed TL-nvSRAM-CIM cell and previous SL-nvSRAM-CIM [12].**

| Circuit Metrics | 6T SRAM | SL-nvSRAM-CIM* | **Proposed TL-nvSRAM-CIM** |
|---|---|---|---|
| Store Energy (fJ) | NA | 360 | 69.2 |
| Restore Energy (fJ) | NA | 15.6 | 8.57 |
| Data Stored per Cell | 1 bit | 18 bits | 240 trits (384 bits) |
| CIM Efficiency (op/fJ) | NA | 0.58 | 0.85 |
| Cell Area (µm$^2$) | 0.75 | 2.33 | 6.35 |
| Storage Density (bit/µm$^2$) | 1.33 | 7.73 | 60.47 |

*Simulated and evaluated in 28nm technology.*

**Table 5: Design specifications of evaluated circuits in 28nm CMOS technology.**

| Component | Parameter | Spec. |
|---|---|---|
| SRAM-CIM/ SL-nvSRAM-CIM | Rows activated | 32 rows |
|  | Energy | 0.11 pJ/col. |
| TL-nvSRAM-CIM | Subarray Size | 256×320 |
|  | Rows activated | 16 rows |
|  | Energy of CIM | 0.096 pJ/$CBL$ |
|  | Energy of restore | 75.2 pJ/array |
| Ternary Encoder | Energy | 13.1 fJ/conversion |
| ADC | Energy | 0.188 pJ |
| Shift & Add | Energy | 0.3360 pJ/5col. |
| Buffer | Energy | 0.042 pJ/bit |
| DRAM | Read Energy [21] | 4.2 pJ/bit |
|  | Read delay | 1 ns |
| ReRAM for storage | Read Energy [22] | 1.63 pJ/bit |
|  | Read delay | 5 ns |

nSnRs with DC-power-free weight-restore operation, the energy consumptions of storing and restoring weights in TL-nvSRAM-CIM cell are reduced by 80.7% and 45.1%, respectively, as compared to the SL-nvSRAM-CIM. By employing the energy-efficient ternary MAC operation, the CIM efficiency of the TL-nvSRAM-CIM is increased by 46.6% compared to SL-nvSRAM-CIM.

In the standard 6T SRAM, only 1 bit data can be stored in each cell. Considering the reliability of circuit under both CMOS and ReRAM variations, the total ReRAM number in the previous SL-nvSRAM-CIM cell is set with 3 groups and 6 SL-ReRAMs per group. In the proposed TL-nvSRAM-CIM cell, there are 4 TL-ReRAM clusters and each cluster contains 60 TL-ReRAMs, without extra area for accommodating ReRAMs. Note that, although TL-nvSRAM-CIM cell has much larger silicon area than SL-nvSRAM-CIM cell, the increased silicon area does not necessarily lead to storage density loss. This is because more TL-ReRAMs can be placed onto the top of the increased silicon area, thanks to the BEOL-compatible nature of ReRAM. The proposed TL-nvSRAM-CIM significantly improves the data capacity of each cell to 384 bits (equivalent to 240 trits), compared to the 18 bits in the previously published SL-nvSRAM-CIM. The storage density is calculated by the ratio of the number of equivalent stored weight bits and the layout area of each cell. As listed in Table 4, the storage density of proposed TL-nvSRAM-CIM cell is significantly improved by 7.8x, compared to SL-nvSRAM-CIM cell.

### 4.3 TL-nvSRAM-CIM Architecture

***Throughput.*** The peak throughput of binary based SL-nvSRAM-CIM and ternary based TL-nvSRAM-CIM is shown in Fig. 9(a). The proposed ternary based TL-nvSRAM-CIM achieves 1.3x enhanced peak throughput compared to the SL-nvSRAM-CIM. Note that, even designed with array size of 256×250, the proposed TC-based TL-nvSRAM-CIM can still achieve similar throughput as SL-nvSRAM-CIM while significantly reducing 21.9% ADCs (from 32 to 25) in a single array.

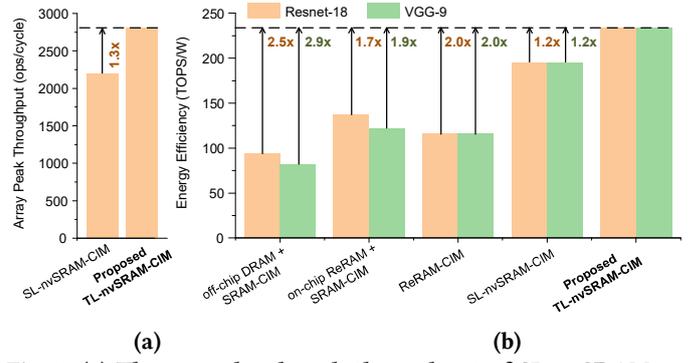

**Fig. 9.** (a) The array-level peak throughput of SL-nvSRAM-CIM and TL-nvSRAM-CIM (normalized to 1b). (b) The evaluation results of the energy efficiency of the proposed TL-nvSRAM-CIM and other baselines.

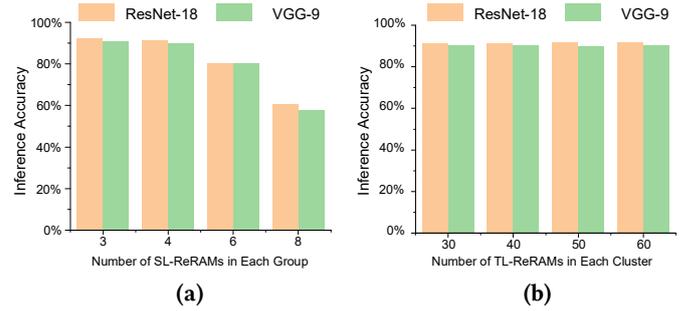

**Fig. 10.** The evaluations of the inference accuracies with retraining for (a) SL-nvSRAM-CIM and (b) TL-nvSRAM-CIM across different ReRAM settings, considering the restore yield loss under both CMOS and ReRAM variations.

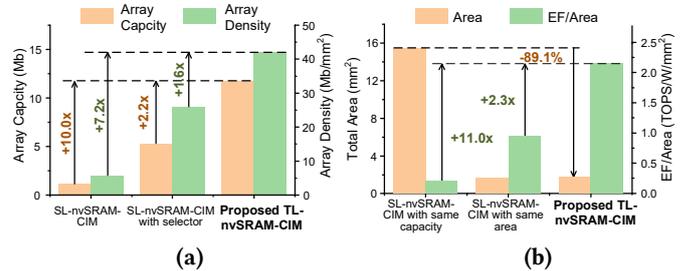

**Fig. 11.** (a) The comparisons of array capacity and array storage density between SL-nvSRAM-CIM and TL-nvSRAM-CIM. (b) The evaluations of the area and energy efficiency per unit area on ResNet-18 (normalized to 8b).

***Energy efficiency.*** The evaluations of energy efficiency for different inference tasks are shown in Fig. 9(b). With all of weights stored in on-chip clustered TL-ReRAMs and the low energy cost of weight restore operation, the energy efficiency of ResNet-18 and VGG-9 inference of the proposed TL-nvSRAM-CIM is improved by 2.5x/2.9x and 1.7x/1.9x, respectively, compared to the traditional SRAM-CIM with weights stored in off-chip DRAMs of baseline-1 and on-chip ReRAMs of baseline-2. Meanwhile, with the energy-efficient SRAM-CIM mechanism, the TL-nvSRAM-CIM enhances the energy efficiency by 2.0x on two NN models, compared to the ReRAM-CIM baseline-3. Moreover, thanks to the 8-bit to 5-trit conversion of inputs and the high-efficient ternary MAC operations, the energy efficiency is enhanced by 1.2x, compared to the SL-nvSRAM-CIM. Note that, the CIM energy of TL-nvSRAM-CIM is reduced compared to SL-nvSRAM-CIM in a tricky way since it is strongly correlated to the parasitics of manually designed layout. To tackle this, we further evaluate the scenario that SL- and TL- nvSRAM-CIM all have the same CIM energy, the proposed TL-nvSRAM-CIM can still achieve 1.15x enhanced energy efficiency.

*Accuracy.* The accuracies of ResNet-18 and VGG-9 on Cifar-10 dataset for previous SL- and proposed TL-nvSRAM-CIM under both CMOS and ReRAM variations are shown in Fig. 10. To achieve the similar accuracy as TL-nvSRAM-CIM, the ReRAM number in each SL-nvSRAM-CIM cell is limited for high restore yield (low bit error rate). Alternatively, the capacity of each TL-nvSRAM-CIM cell is significantly enhanced to 240 trits per cell by employing the cluster-nSnRs. Thanks to the reliable DC-power free restore operations, the inference accuracy can be well maintained regardless of the number of ReRAMs pers cluster. However, the previous SL-nvSRAM-CIM has constraint ReRAM numbers per group to avoid the accuracy loss, showing poor scalability for an increased number of ReRAMs.

*Capacity/density benefits.* The ablation study for the array capacity and array storage density of SL-nvSRAM-CIM and TL-nvSRAM-CIM is shown in Fig. 11(a). For fair comparison, the number of clusters in TL-nvSRAM-CIM is assumed to be 3 in the ablation study. The evaluations of array area include memory array and peripheries such as ADCs, shift & adder, and ternary encoders. Since the selector based select scheme can also be adopted to SL-nvSRAM-CIM, the number of ReRAMs contained in a single group can be increased to 18 which leads to 4.5x enhanced array capacity and storage density compared to [12]. TL-nvSRAM-CIM further adopts the multi-level (ML) property of ReRAM, leading to a total 10.0x/7.2x enhanced array capacity/storage density. The evaluations of the area overhead and energy efficiency per unit area in the inference process of ResNet-18 are shown in Fig. 11(b). To accommodate the whole ResNet-18 model, 76 SL-nvSRAM-CIM subarrays or 6 TL-nvSRAM-CIM subarrays are needed. Therefore, 89.1% of area is saved by the proposed TL-nvSRAM-CIM, and the energy efficiency per unit area is enhanced by 11.0x, compared to the SL-nvSRAM-CIM. However, if occupying the same area as TL-nvSRAM-CIM, only part of NN weights can be stored in SL-ReRAMs of previous SL-nvSRAM-CIM due to the limited on-chip storage capacity. Extra data transfer is still required for loading remaining weights from off-chip memories, which degrades the energy efficiency of previous SL-nvSRAM-CIM. Thus, assuming with the same area, the proposed TL-nvSRAM-CIM still significantly enhances the energy efficiency per unit area by 2.3x, compared to the SL-nvSRAM-CIM.

*Discussion.* Rather than solely increasing the data representation level of SL-ReRAMs, our work introduces the following key novelties: 1) We have designed a new topology for nvSRAM circuitry that can reliably support backup and recovery of ternary data without requiring DC-power. This new circuitry enables excellent scalability with an increased number of TL-ReRAMs, maximizing the high-density benefits of ReRAM. 2) We have developed a novel ternary-based SRAM-CIM mechanism with a differential computing scheme that tightly integrates the TL-nvSRAM circuit and the data representations of TL-ReRAM. This approach delivers superior throughput and energy efficiency compared to binary CIM. Furthermore, we have evaluated the impact of CMOS and TL-ReRAM variations on specific inference tasks, demonstrating the robustness of TL-nvSRAM-CIM. While this work does not primarily focus on auxiliary techniques [7][8] that control the programming process of TL-ReRAMs, as the programming phase only occurs during initialization, these techniques can be further explored in future work with the TL-nvSRAM-CIM scheme to enhance reliability.

## 5 CONCLUSION

We propose a new TL-nvSRAM-CIM circuit and scheme that offer exceptional storage density and energy efficiency. By utilizing the high-density TL-nvSRAM-CIM array, we achieve a 7.8x increase in storage density compared to the previous SL-nvSRAM-CIM. Furthermore, the power consumption of weight loading is significantly reduced through the DC-free restore mechanism. Our evaluation results demonstrate that the TL-nvSRAM-CIM scheme achieves up to 2.9x greater energy efficiency in neural network inference compared to the traditional CIM method with memory access. Additionally, the proposed TL-nvSRAM-CIM enhances energy efficiency per unit area by 11.0x, with 89.1% of the area saved. Overall, the TL-nvSRAM-CIM circuit and scheme offer remarkable benefits in terms of storage density and energy efficiency.

## ACKNOWLEDGMENTS

This paper is supported in part by National Natural Science Foundation of China (#62174110, #U21B2030, #92264204), National Key R&D Program of China (#2021YFA0717400, #2019YFA0706100), Natural Science Foundation of Shanghai (#23ZR1433200), and BirenTech Research.